# Tunable spider-web inspired hybrid labyrinthine acoustic metamaterials for low-frequency sound control

A.O. Krushynska[1(*)], F. Bosia[1], M. Miniaci[2], N.M. Pugno[3,4,5(**)]

[1]Department of Physics and Nanostructured Interfaces and Surfaces Inter-departmental Centre, University of Torino, Via Pietro Giuria 1, 10125 Torino, Italy

[2]University of Le Havre, Laboratoire Ondes et Milieux Complexes, UMR CNRS 6294, 75 Rue Bellot, 76600 Le Havre, France

[3]Laboratory of Bio-Inspired and Graphene Nanomechanics, Department of Civil, Environmental and Mechanical Engineering, University of Trento, Via Mesiano 77, 38123 Trento, Italy

[4]School of Engineering and Materials Science, Queen Mary University of London, Mile End Road, London E1 4NS, United Kingdom;

[5]Ket Labs, Italian Space Agency, Via del Politecnico snc, 00133 Rome, Italy

Corresponding authors: (*) akrushynska@unito.it and (**) nicola.pugno@unitn.it

**Abstract**

Attenuating low-frequency sound remains a challenge, despite many advances in this direction. Recently developed acoustic metamaterials enable efficient subwavelength wave manipulation and attenuation due to exotic effects such as unusually high reflectivity, negative refraction or cloaking. In particular, labyrinthine acoustic metamaterials can provide broadband sound reduction and exhibit extremely high effective refractive index values due to their characteristic topological architecture. In this paper, we design a novel labyrinthine metamaterial with hybrid characteristics compared to previously proposed structures, by exploiting a spider web-inspired configuration. The developed metamaterial structure is characterized by additional tunability of the frequencies at which band gaps or negative group velocity modes occur, thus enabling versatility in the functionalities of the resulting structures. Time transient simulations demonstrate the effectiveness of the proposed metamaterials in manipulating wave fields in terms of transmission/reflection coefficients, amplitude attenuation and time delay properties in broadband frequency ranges. Results could find applications in the development of practical lightweight acoustic shielding structures with enhanced broadband wave-reflecting performance.

**Keywords:** Acoustic metamaterials, labyrinthine acoustic metamaterials, low-frequency sound attenuation, wave dispersion, hybrid properties, bioinspired structures, finite element method

1. Introduction

Manipulating low-frequency sound remains a challenging task for scientists and engineers, despite a vast amount of research in this field. Efficient shielding of sound by means of conventional natural or artificial materials with low refractive indexes requires the construction of impractically thick and heavy structures that are not economically sustainable [1]. Recently developed acoustic metamaterials offer a new approach for manipulating low-frequency sound waves thanks to their extraordinary functionality originating from their structure rather than from their material constituents.

Acoustic metamaterials are engineered composites with unusual effective characteristics, e.g. negative bulk modulus and/or mass density, zero or negative effective refractive index, etc. These metamaterials can formally be divided into locally resonant metamaterials with basic units incorporating resonators [2, 3] and other metastructures described by effective medium theories [4, 5, 6]. The latter comprise so-called "labyrinthine" or "space coiling" metamaterials with a geometry-based mechanism for controlling subwavelength acoustic waves [7, 8, 9, 10, 11].

The advanced functionalities provided by labyrinthine acoustic metamaterials (LAMMs) are based on the concept of creating a network of internal channels (labyrinth) in non-resonant unit cells in order to extend wave propagation paths. This results in the substantial reduction of the effective wave propagation velocity with respect to a homogeneous medium. As a consequence, extraordinary characteristics can be obtained, such as extremely high refractive indexes, negative refraction and "double negativity" (i.e. simultaneous negativity of the effective density and bulk modulus) in effective material properties [10, 11]. The space-coiling approach opens a simple and reliable way for designing versatile, easy-to-fabricate acoustic metamaterials with isotropic response, which are capable of controlling waves at low, so-called "deep subwavelength" frequencies [8]. This is a particularly desirable characteristic when large wavelength vibrational phenomena need to be addressed, such as in the case of large scale mechanical metamaterials for seismic shielding [12].

The first zig-zag type labyrinthine metamaterial was proposed by Liang and Li [7]. Besides a very high refractive index, which is rarely found in natural and engineered materials, this metamaterial was shown to exhibit unique characteristics such as negative refraction, conical dispersion, and near-zero or extreme (positive or negative) effective properties. This structure and other similar ones were later investigated experimentally for acoustic [8, 9] and electromagnetic [8] waves at kHz and GHz frequencies, respectively. A corresponding three-dimensional metastructure was also reported and tested for acoustic waves in the kHz regime [10]. The extreme effective properties of zig-zag-type labyrinthine metastructures result in a high impedance mismatch with air. To address this issue, several spriral-like metamaterial configurations with tapered channels and apertures were proposed [13]. The ability of labyrinthine structures to reduce the amplitude of propagating waves due to multiple wave reflections within labyrinthine channels was extended by developing ultra-sparse highly-reflecting acoustic absorbers for low-frequency sound [11, 14]. Their functionality is based on intense artificial Mie resonances in the low-frequency range, allowing almost perfect reflection of low-frequency airborne sound waves [11]. These resonances enable the generation of subwavelength band gaps, but eliminate conical dispersion and bands with negative group velocity from the dispersion characteristics of the metastructure.

In this paper, we propose *hybrid* labyrinthine acoustic metamaterial structures, simultaneously achieving subwavelength band gaps and negative group velocity in low-frequency dispersion bands. The hybridization is achieved by modifying the labyrinthine structure, adding a square external frame to circular-shaped curved channels. This feature also results in a non-uniform width of the curved channels and the creation of additional cavities at the edges of the unit cell. We show that the presence of the cavities determines the wave propagation characteristics of these metamaterials and enables the activation or removal of subwavelength band gaps by varying the cavity size. This tunability is important for engineering applications of LAMMs, enabling their practical utilization, and could potentially lead to new application areas. Also, studies on the dependence of labyrinthine metamaterials on geometrical parameters in view of structural optimization have thus far not been undertaken. Thus, we have analyzed in detail the influence of different types of periodic and non-periodic inhomogeneities in the channel width on the wave manipulation performance of the proposed structures.

The overall design of the developed metamaterials as well as the introduced variations are inspired by the geometry of natural spider webs typically consisting of a bearing frame and two spirals separated by a cavity [15, 16, 17, 18, 19, 20]. To simplify the analysis, we replace a polygonal spider-web frame and two spirals with a square frame and a number of circular geometries. This design is used for the arrangement of solid walls governing wave propagation. Recent application of spider web-inspired geometries as locally-resonant acoustic metamaterials has revealed several advantages of this architecture for low-frequency elastic wave manipulation due to the interplay between the frame and circular resonators [21]. In this work, we demonstrate that adding the spider web-inspired frame to a circular "labyrinth" of curved channels enables to improve the highly reflective performance of the latter and to achieve tunability of wave manipulation characteristics by exploiting the additional degrees of freedom.

## 2. Model and Methods

### 2.1. Structure description

The designed LAMM configurations consist of a square external frame and a circular "labyrinth" divided into either four or eight independent curved channels connected at the center. Schematic diagrams in Figs. 1a,b show the unit-cell cross sections with thin elements indicating solid walls (depicted in blue) surrounded by air (depicted in violet). We denote these structures as *hybrid* LAMMs (HLAMMs) to distinguish them from conventional LAMMs, which have been analyzed previously [11]. Form a geometrical point of view, our structures differ from previous designs for the presence of edge cavities. From a physical point of view, these cavities govern the wave propagation characteristics of the designed HLAMM and significantly influence the wave manipulation properties at low frequencies, as discussed below. The structures we study include the limiting case of a square frame completely filled by curved channels, i.e. with the size of edge cavities tending to zero, which we shall refer to as "square HLAMM" for convenience. An example of this type of structure with 8 curved channels is shown Fig. 1c.

HLAMM geometries can be defined by seven parameters: the lattice constant $a$; the thickness of solid walls $d$; the length of side walls $l$; the radius of the internal cavity $r$; the width of a passage between channels $h$, the curling number $N$ (or the number of circles used to create the curved channels) and the external radius $r_i$ ($i = 1, ..., N$) of curved walls (these parameters are indicated in Fig. 1a). We consider $a = 0.9$ m, $d = 0.01$ m, $l = 0.73$ m, $r = 0.05$ m, $h = 0.05$ m, $N = 7$, $r_i = 0.05 \cdot (i + 1)$, $i = 1, ...,7$. The walls are considered made of aluminum with Young's modulus $E = 70$ GPa, Poisson's ratio $\nu = 0.33$, and mass density $\rho = 2700$ kg/m³. Standard values for air, i.e. mass density $\rho_0 = 1.225$ kg/m³, atmospheric pressure of $10^5$ Pa, and speed of sound $c = 343$ m/s, are used.

In terms of a geometric acoustics representation, sound waves propagate within the unit cells along the curved channels as indicated by the thick solid line in Fig. 1b instead of along the straight dashed line linking the external point $P_1$ to the unit cell center $P_2$. As a result, the wave path is increased by a factor of approximately 4 and 2.4 for four and eight-channel unit cells, respectively. This factor is known as coiling coefficient $\eta$ and is evaluated as the ratio between the approximate total length of the path and the unit cell pitch [13]. For the eight- and four-channel unit cells with edge cavities, the coiling coefficient equals to $\eta^{(8)} = 3.4$ and $\eta^{(4)} = 5.5$, respectively. A rough estimation of the metamaterial refractive index as a function of frequency can then be obtained as $n_{eff} = \eta - 2\pi c/\omega a$ [9].

*2.2. Calculation of dispersion spectra*

Dispersion properties of the HLAMMs are numerically evaluated by means of finite-element simulations using the Pressure Acoustics module of COMSOL Multiphysics. We consider propagation of plane pressure waves $p = p_0(x)e^{i\omega t}$ of angular frequency $\omega$ governed by the Helmholtz equation:

$$\nabla \cdot \left(-\frac{1}{\rho_0} \nabla p_0\right) - \frac{\omega^2}{\rho_0 c^2} p_0 = 0 \qquad (1)$$

in an infinite periodic HLAMM medium modeled by applying Floquet-Bloch periodic conditions at the unit cell boundaries:

$$p_0(x + a) = p_0(x)e^{-ka}. \qquad (2)$$

Given the unit cell length $a = 0.9$ m and rather small curling number $N = 7$, the sound pressure field may be regarded as lossless in the sense that thermo-viscous boundary layers accompanying wave propagating can be neglected [11, 8]. At the solid walls, we prescribe sound-hard boundary conditions that imply vanishing normal sound velocity, i.e. $\partial p_0/\partial n = 0$, where $n$ indicates an external normal to a wall. This is justified by an impedance value of aluminum four orders larger than that of air, as has also been experimentally verified [10, 22].

*2.3. Transmission analysis*

The ability of the proposed HLAMMs to block the propagation of low-frequency acoustic waves can be confirmed by analyzing their transmission characteristics. In particular, we examine plane pressure waves propagating along the horizontal direction ($\Gamma$-$X$) through a varying number of closely located unit cells (the center-to-center distance between unit cells is $a$). The simulations are performed in the frequency domain using COMSOL Multiphysics.

*2.4. Transient analysis*

To estimate wave manipulation capabilities of the developed HLAMMs at low frequencies, we numerically simulate a pulse propagation through a set of metamaterial unit cells and evaluate the amount of transmitted and reflected wave energy. The analysis is performed for the eight-channel HLAMM unit cells and the corresponding conventional LAMMs.

The time-dependent acoustic pressure field $p(\boldsymbol{x}, t)$ is governed by the scalar wave equation,

$$\nabla \cdot \left(-\frac{1}{\rho}\nabla p\right) + \frac{1}{\rho c^2}\frac{\partial^2 p}{\partial t^2} = S(t). \qquad (3)$$

Source waves are unity-amplitude Gaussian modulated sinusoidal pulses of in the form of

$$S(t) = \sin(2\pi f_c t)\exp(-t^2/2\sigma^2) \qquad (4)$$

with $f_c$ denoting a carrier frequency, $\sigma$ being the deviation of a pulse, and the pulse width $2\pi\sigma$.

The simulation model is shown in Fig. 2 and consists of a number of closely-located unit cells (or single unit cell) at a center-to-center distance of $a$. At the top and bottom domain boundaries, periodic continuity conditions are applied to simulate an infinite medium in the vertical direction. On the right boundary, the plane wave radiation condition is used in order to minimize spurious reflections of normally incident waves. Uniformly distributed pressure (Eq. (4)) is applied on the left boundary to simulate plane acoustic waves, such as those launched by a loudspeaker located at a large distance from the metamaterial region (similar, e.g. to the experiments reported in Ref. [10]). The acquisition point is chosen at a distance $a$ on the right side from the unit cells. The simulations are numerically performed by means of the commercial COMSOL Multiphysics by neglecting any type of losses due to the chosen sizes of the metastructures [11].

## 3. Results
### 3.1. Dispersion characteristics

Figure 3a shows a dispersion diagram for a HLAMM unit cell with eight curved channels. The horizontal axis indicates the values of wave vector $\boldsymbol{k}$ along the borders ($\Gamma$-X-M) of the triangular irreducible Brillouin zone for a square periodic lattice [23], while the vertical axis shows normalized values of frequencies, $\Omega = \omega a/2\pi c = fa/c$. In this work, we focus our attention on the subwavelength frequency range; and thus, the maximum value of the considered frequencies is restricted to $\Omega = 0.525$ ($f = 200$ Hz), which corresponds to the case when the wavelength of an airborne wave is about two times larger than the unit cell pitch. The distinctive characteristics of the spectrum are the presence of bands with negative group velocity, three subwavelength band gaps and several flat bands with almost zero group velocity.

Figures 3b and 3c represent dispersion diagrams for a conventional LAMM and a square HLAMM, respectively. The dispersion spectrum for the conventional LAMM (Fig. 3b) includes the presence of low-frequency band gaps while that of the square HLAMM (Fig. 3c) displays the presence of several dispersion bands with negative group velocity. The wave dispersion in the latter case resembles that for zig-zag-type labyrinthine metamaterials [10, 7, 14, 8] which display negligible sub-wavelength band gaps and conical dispersion, also observed in Fig. 3c.

When comparing Figs. 3b and 3c with Fig. 3a, it is apparent that the HLAMM has hybrid dispersion characteristics between those of the conventional LAMM and a square HLAMMs. The proposed HLAMM exhibits low-frequency band gaps similar to the conventional LAMM but also has dispersion bands with a negative group velocity like the square HLAMM. For the same geometric parameters of the unit cells, the band gaps for the HLAMMs (Fig. 3a) are located at frequencies about 1.5 times lower than the conventional LAMM (Fig. 3b). Thus, due to the scalability of dispersion properties, unit cells with $a = 0.09$ cm should allow the manipulation of waves with frequencies down to 400-600 Hz, while maintaining a reduced total weight compared to most locally resonant structures. By varying the size of the edge cavities, the band gaps can be shifted in frequency or even completely closed, when they are filled by curved channels (such as in Fig. 3c).

The dispersion spectra in Figs. 3a and 3c highlight several localized modes characterized by pressure distributions similar to the artificial Mie monopole and multipole resonances first found for the conventional LAMMs [11]. The corresponding mode shapes are shown as insets to the dispersion spectra in Fig. 3. It is well-known that monopole Mie resonances are directly related to negative values of an effective bulk modulus, which occur at frequencies above this mode [24]. For the conventional LAMM and the HLAMM, the normalized frequencies of the monopole Mie resonance are 0.186 and 0.168, respectively. They form a lower and upper bound of the first (lowest) band gap, respectively. The dipole Mie resonance is associated with negative effective mass density occurring below the resonance frequency [24]. In Figs. 3a and 3b, this mode is found at $fa/c = 0.331$ and 0.379, respectively, and forms the upper bound of the second band gap. If frequency ranges with negative values of an effective bulk modulus and effective mass density are overlapped, double-negativity phenomena can be achieved [9, 8] (beyond the scope of this paper).

For the square HLAMM, pressure distributions at normalized frequencies 0.123 and 0.254 also resemble those for artificial monopole and dipole resonances. However, the band gaps around these frequencies are of negligible size. A possible reason for this is insufficient vibration localization within the unit cell with negligible zero-pressure zones. Thus, wave energy is not sufficiently trapped, and is allowed to propagate. These results are also confirmed by transmission analysis, discussed below.

*3.2. Transmission analysis.*

The spectra in Figs 3d, 3e and 3f show the magnitude of normalized absolute pressure, $log \int_A |p_{tr}/p_{in}| dA$ (with subscripts indicating transmitted and incident pressure, and $A$ designating the area of a single unit cell), versus the normalized frequency $fa/c$. High transmission losses are found at the band gap frequencies predicted by the dispersion analysis. The slight discrepancies are attributed to the finite size of the model for the transmission simulations.

Clearly, the magnitude of the transmission reduction depends on the number of the involved unit cells. For example, five unit cells of the conventional LAMM ensure significant transmission losses within the second band gap (Fig. 3e), while the same number of unit cells of the HLAMM (Fig. 3d) provide a less efficient attenuation, by about 3 orders of magnitude. This is due to insufficient vibration localization inside the metastructure, as discussed above. Transmission losses can be increased proportionally to the number of the unit cells, as shown in Fig. 3d. Taking this into account, a practical configuration of a compact-size HLAMM can be achieved by using an approach proposed in Ref. [10] that implies stacking unit cells one above the other.

The transmission spectrum for the square HLAMM (Fig. 3f) indicates the absence of transmission losses regardless of the number of involved unit cells. This behavior is again similar to that of the zig-zag-type labyrinthine metamaterials [7, 10]. Multiple small dips in transmission seen in Fig. 3f and similar dips below the first band gap in Fig. 3d correspond to zero pressure points at the mode nodes . For example, the dip at about $f = 5$ Hz in Fig. 3d appears to be due to the first spatial node of a mode at a wavelength $\lambda = c/f = 68.60$ m. This node is located at $\lambda/4 = 17.15$ m, i.e. approximately the distance from the wave source to the acquisition point through 5 unit cells of the metamaterial (for a single unit cell, $\eta a\sqrt{2} \approx 3$m) and three homogeneous air unit cells (2.7 m long) also present in the model.

Finally, we note that the hybrid features of the proposed HLAMM are also found for a four-channel configuration, the dispersion and transmission characteristics of which are given in Fig. 4. As this metamaterial has a higher coiling coefficient, the band gaps are shifted to even lower frequencies (compared to those in Fig. 3a). At the same time, frequency regions with negative group velocity are reduced. Thus, by decreasing a number of the curved channels from 8 to 4, the sound-blocking abilities of the proposed metamaterials can be shifted lower frequencies.

*3.3. Tunability of Mie resonances and band gap frequencies.*

As demonstrated in the previous section, artificial Mie resonances govern the subwavelength band gap location. Hence, by changing their frequencies, one can tune band gaps to desired ranges. In previous work, it was suggested to tune the Mie resonances by changing the size and refractive index of the unit cell [11]. In this work, we propose an alternative approach, based on the introduction of inhomogeneities in the channel geometries.

The introduction of these inhomogeneities is inspired by the architecture of a spider web [15, 16]. Fig. 5a shows the structure of a typical spider web built by a garden spider (*Araneus diadematus*). One can observe variable distances between circumferential elements and missing links (defects) in the circumferential silks, thus forming "internal cavities". These types of asymmetries or irregularities can be introduced into the HLAMM geometry to influence its wave manipulation abilities.

Figures 5b-5e present several modifications of the original HLAMM unit cells. Figure 5b shows an 8-channel unit cell from which a curved wall of radius $r_6$ has been removed, resulting in the formation of an internal cavity between the curved channels, leading to a variation in the coiling coefficient and the effective refractive index of the HLAMM. The presence of the cavity also results in a shift the first band gap to higher frequencies and the closing of the second and third band gaps (the corresponding dispersion diagram is not shown). Instead of these band gaps, two dispersion cones are formed similar to those in Fig. 3c. The frequency of the monopole is shifted from 0.168 to 0.194 (+16%), while the dipole resonance remains unchanged. From a physical point of view, the disappearance of the band gaps is attributed to a detrimental effect of the introduced cavity on the vibration localization at the multipole Mie resonances. This effect also leads to an increase in the monopole resonance frequency, since a conventional LAMM of smaller length supports eigenmodes of smaller wavelength. If the cavity occurs between the walls of radii $r_4$ and $r_6$, the dispersion spectrum has two band gaps between 0.144 and 0.173, 0.412 and 0.543 with monopole and dipole Mie resonances at 0.173 and 0.320, respectively. Thus, the second band gap is closed for the same reason as in the previous case. The absence of any wall of radius $r_i$ with $i$ =1-4 does not significantly affect the dispersion properties of the HLAMM: the dispersion curves have similar trends to those in Fig. 3a, the three band gaps are preserved, and the frequencies of the Mie resonances are only slightly shifted from the values for the regular unit cell configuration (Fig. 1a). Therefore, the incorporation of a cavity in the curved channels when eliminating a large-radius curved wall enables to reduce the highly reflective properties of the unit cell attributed to multipole Mie resonances, and to simultaneously introduce conical dispersion by preserving the first subwavelength band gap governed by the monopole resonance.

Figure 5c provides another example of the incorporation of an internal cavity between curved channels, with circular channel radii indicated in Table 1.The channel width in the two regions is different, replicating the structure of a spider web (Fig. 5a). Since we aim to reproduce a spider-web architecture, the cavity (in all the considered configurations) is located closer to the unit cell center, between the walls of radii $r_3$ and $r_4$. First, we consider "Case 1" in which the width of the channel part external (or internal) to the cavity equals $0.045a$ (or $0.03a$). These parameters allow to approximately preserve the coiling coefficient equal to that of the regular configuration (Fig. 1a). The dispersion diagram for Case 1 is the same as the one in Fig. 3a up to the normalized frequency 0.210. At higher frequencies, most of the bands are shifted to lower frequencies as compared to the regular case, with a shift in the dipole resonance frequency from 0.331 (for the regular unit cell) to 0.311 (a -6% shift). The size of the second band gap is almost unchanged, while the third band gap is increased by about 1.5 times. Next, we consider "Case 2" in which part external to the cavity has the same channel width as the regular configuration, i.e. $0.05a$, and the internal part a smaller width ($0.02a$). The dispersion diagram is again similar to that in Fig. 3a with bands shifted to lower or higher frequencies. In particular, the first band gap is two times smaller in size, as the lowest band is shifted to higher frequencies, though the monopole frequency is unchanged. The second bad gap almost disappears, since the dipole resonance is shifted down to a normalized frequency of 0.257. The third band gap (governed by the quadrupole Mie resonance) is enlarged and spans from a normalized frequency of 0.346 to 0.460. If the cavity size is further increased by reducing the width of the channels in the external part to $0.035a$ ("Case 3"), the third band gap is further enlarged and becomes 3.7 times larger than that of the regular configuration. In this case, the monopole and dipole frequencies are at normalized frequencies of 0.129 and 0.231, respectively.

Thus, an internal cavity and varying width of the channel parts does not modify dispersion characteristics of the labyrinthine metamaterials, since subwavelength band gaps and the general structure of the dispersion spectrum are preserved. However, the frequencies of band gaps and the Mie resonances can be efficiently tuned. In general, to obtain a non-negligible shift of the monopole frequency, it is necessary to introduce a large cavity. This is because there are small changes in the coiling coefficient values and external sizes of the curved channels $r_1$ and $r_7$, which determine the frequency of the monopole resonance [11], are fixed. However, the dipole (and quadrupole) resonances can easily be tuned, and thus, the second and third band gap frequencies, since the introduction of a cavity influences the location of the vibration maxima in the corresponding mode shapes. In particular, a combination of a large cavity and thin curved channels (as in Case 3) enables the achievement of an extremely wide third band gap. This feature can be effectively exploited to design broadband low-frequency sound absorbers.

Finally, we consider unit cells with a periodically varying width of the curved channels, shown in Figs. 5d-e. These configurations are not spider web-inspired, but due to a limited number of parameters, allow a simplified analysis. These configurations can be described by the following values of external radii: $r_i = 0.05(i + j) + 0.02$ for $i = 2,4,6$ and $j = 0,1, ... ,8$. Figures 6a and 6b show normalized frequencies of subwavelength band gaps and the monopole and dipole Mie resonances for the eight- and four-curved channel metamaterial configurations, respectively, as a function of the parameter $d_1/2d$, where the $d_1$ dimension is schematically shown in the inset on the left. For a regular configuration (one with curved channels of constant width), $d_1/2d = 50\%$. The monopole and dipole resonances can be shifted up to 40 % by exploiting a periodically varying channel width. Maximum variations occur for $d_1 < 50\%$. Again, as for the spider-web inspired unit cells, the frequency of the monopole resonance varies less as compared to the dipole. The band gap frequencies vary as a function of the channel width. The first band gap is, in general, shifted to lower frequencies compared to the regular configurations, and can even be closed, when $d_1/2d \approx 20\%$. The second band gap also tends to lower frequencies and becomes wider for $d_1/2d \geq 60\%$. Thus, the exploitation of a periodically varying channel width enables to access a challenging low-frequency range while preserving structural sizes. This feature can also be of importance for practical applications of HLAMMs as low-frequency sound reflectors.

*3.4.Transient analysis*

Figure 7 shows calculated transmitted (solid blue line) and reflected (dashed blue line) signals for the HLAMM composed of five unit cells and for the conventional LAMM. Note that the reflected signal is estimated at a point beyond the metamaterial structure at a distance $a/2$ and also includes the incident signal. Incident pulses of different carrier frequencies $f_c$ are indicated in red. They are calculated for the reference configuration, in which the metamaterial region in Fig. 7 is removed. The absence of reflected waves in the latter case proves the reliability of the radiation condition at the right boundary.

As expected, for every carrier frequency, the signal transmitted through the metamaterial part is shifted to a later time, due to the lower effective sound propagation velocity in the metamaterial.

Also, the signals are substantially attenuated. The level of attenuation depends on the carrier frequency and ranges from 25% to 50% for frequencies outside the band gaps and from 60% to 70% within the band gaps. Since viscoelastic losses are not accounted for in the model, this means that a pulse is "trapped" by the metamaterial structure. After multiple reflections, its energy is gradually released and appears as tail pulses to the initial transmitted pulse.

Outside the band gaps, the signal preserves its shape and is followed by a series of self-repeating patterns of a gradually decreasing amplitude. The small magnitude of the reflected waves in the lower subfigure of Fig. 7a proves that most of the incoming energy is transmitted decomposed into several pulses of smaller amplitude and separated in time.

Within the band gaps, on the contrary, the transmitted signal does not preserve its shape due to the peculiar dynamics of the monopole Mie resonance governing the band gap generation mechanism [11]. The influence of this resonance can clearly be seen by analyzing the reflected signals. In the case of the conventional LAMM (Fig. 7b), the magnitude of the reflected signal in the band gap is comparable to that for the pulse with the carrier frequency outside the band gap. For the HLAMM (Fig. 7a), instead, the reflected signal is much stronger, indicating improved reflectivity of the structure achieved due to the presence of the external frame.

**Conclusions**

In this paper, we have proposed labyrinthine metamaterials with hybrid properties with respect to those analyzed in the literature thus far, with attractive properties for low-frequency sound-wave manipulation, drawing inspiration from the architecture of a spider web. The developed metastructures are characterized by the simultaneous existence of monopole and multipole Mie resonances and related band gaps at deep subwavelength scales, and simultaneously by several negative slope bands in a wide frequency rage. The hybridization of dispersion properties is achieved by adding edge cavities to circular curved channels. The key advantage of this design is the versatility in controlling wave manipulation performance, e.g. activation/removal of band gaps by changing the unit cell topological organization, such as the number and width of the curved channels, as well as by introducing cavities externally at the unit cell corners or internally between the channels. The effects of inhomogeneities in the channel width on the wave manipulation performance have been thoroughly analyzed. We have found that the asymmetries incorporated into the HLAMM design, typical for natural webs, enable the tuning of the Mie resonances to even lower frequencies and the extension of the band gap size up to almost four times while preserving structural dimensions. Additionally, we have demonstrated that the proposed metamaterials efficiently attenuate sound waves in a broadband frequency range with different dynamics within and outside the band gap frequencies. Within the band gaps, the highly reflective performance governed by Mie resonances is shown to improve with the interplay of the dynamics of the circular curved channels with that of the square frame. These findings can be useful for applications such as efficient noise-shielding structures with flexible and wideband tunability of the operating frequency ranges in the entire spectrum of low-frequency sound.

**Acknowledgements**

A.O.K. has received funding from the European Union's 7[th] Framework programme for research and innovation under the Marie Skłodowska-Curie Grant Agreement No. 609402-2020 researchers: Train to Move (T2M). M.M. has received funding from the European Union's Horizon 2020 research and innovation programme under the Marie Skłodowska-Curie Grant Agreement No. 658483. N.M.P. is supported by the European Research Council PoC grant 2015 SILKENE No. 693670, EU FETPROACTIVE grant 732344 "NEUROFIBRES", and by the European Commission under the Graphene Flagship (WP14 ''Polymer Nanocomposites'', No. 604391). F.B. is supported EU FETPROACTIVE grant 732344 "NEUROFIBRES". Photo credit: Dr. Dmitry Krushinsky.

# Figures

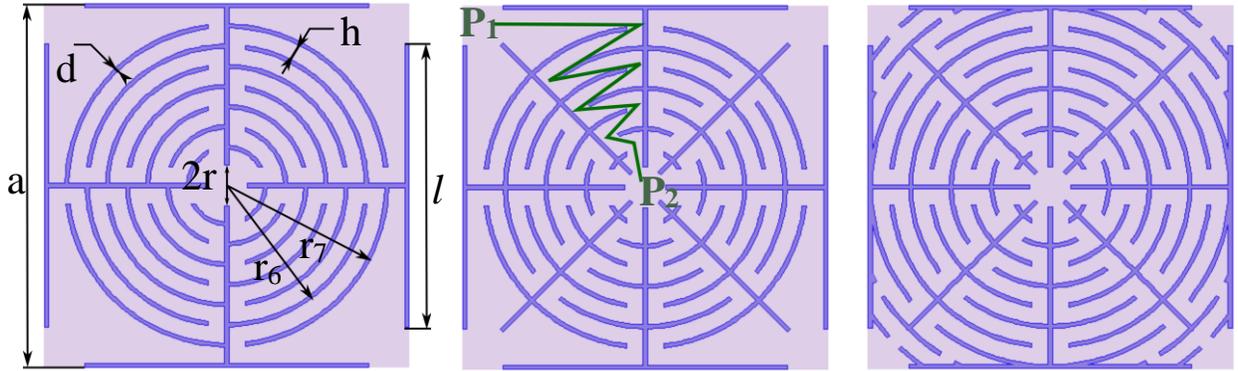

*Figure 1. Unit cells of HLAMMs comprising four (a) and eight (b,c) curved channels connected at the unit cell center. In unit cell (c) (referred to as a "square HLAMM") the curved channels extend to completely fill the edge cavities. Solid aluminum walls are represented in blue, air in violet. The unit cell geometry is determined by 7 parameters indicated in a). Within a unit cell, sound waves propagate from point P1 to point P2 along the curved path indicated in b) by the green line.*

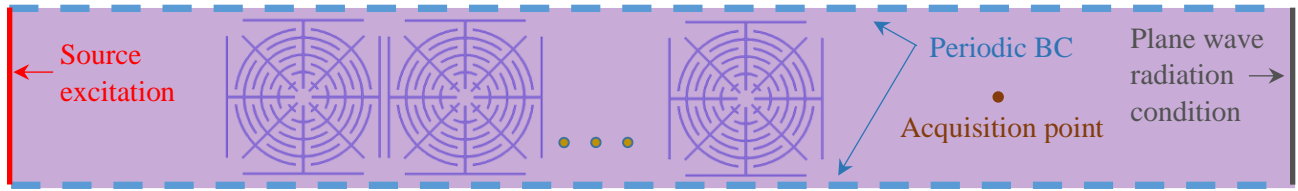

*Figure 2. Model for transient analysis of sound pressure waves propagating thought a set of labyrinthine metamaterial unit cells. A Gaussian modulated sinusoidal pulse is excited on the left boundary, and the transmitted signal is recorded at the acquisition point. Periodic boundary conditions reproduce an infinite medium in the vertical direction, while plane wave radiation condition eliminates signal reflections from the right boundary.*

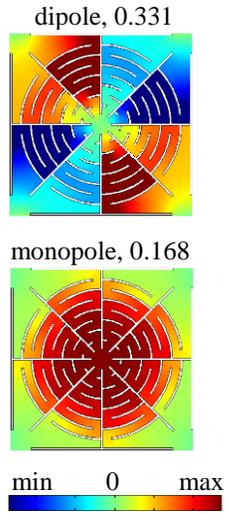
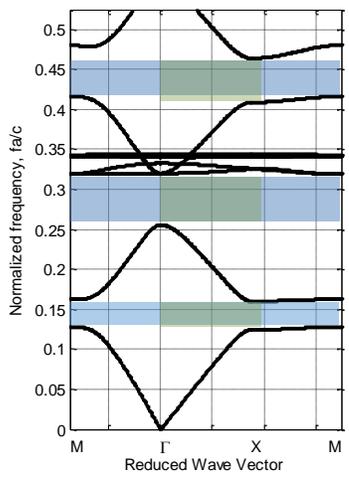
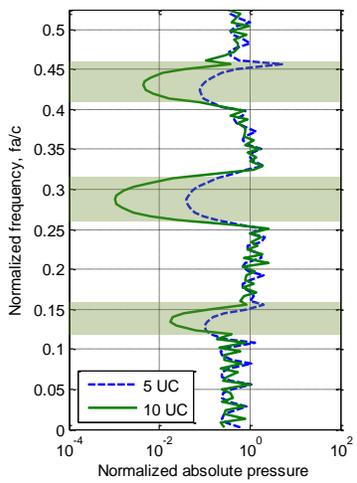

a                           d

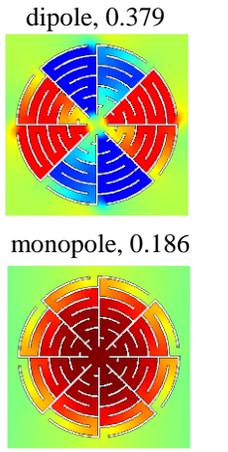
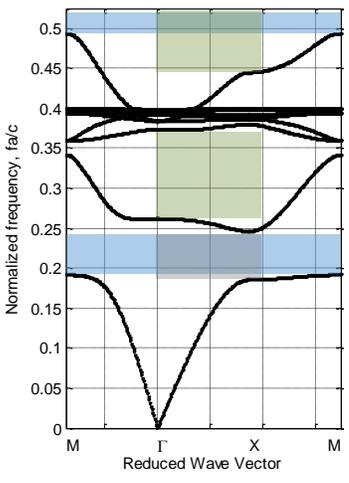
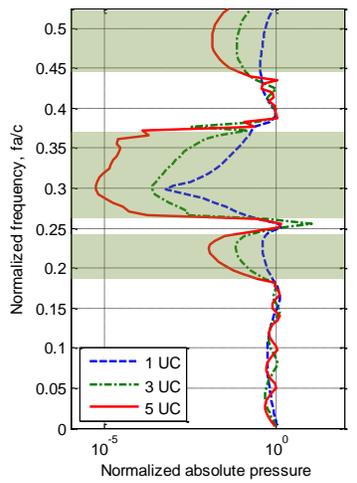

b                           e

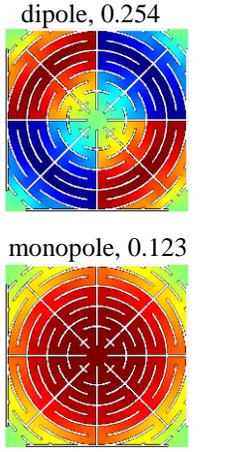
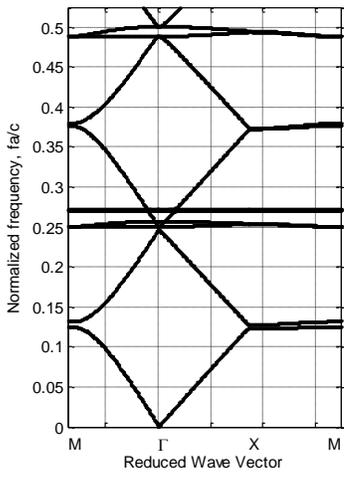
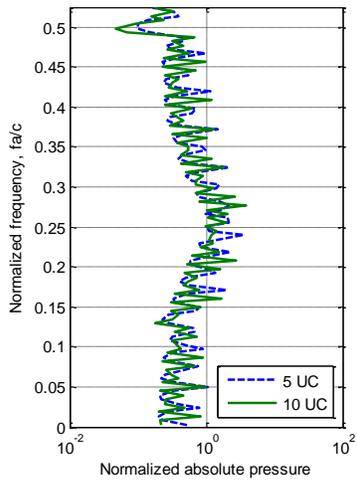

c                           f

*Figure 3. Dispersion and transmission spectra for HLAMMs (a,d), for conventional LAMMs (b,e) and for "square" HLAMMs (c,f). The insets on the left indicate the normalized frequencies and pressure distributions for the artificial monopole and dipole Mie resonances. Omnidirectional band gaps are shaded in blue, while partial band gaps for the ΓX direction are depicted in green. The legends for the transmission spectra indicate the number of unit cells used in the model.*

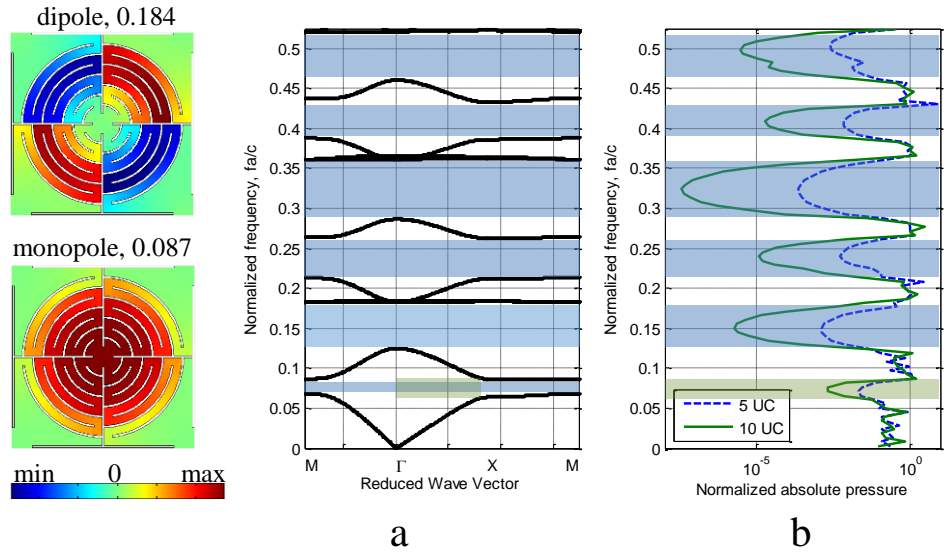

*Figure 4. Dispersion (a) and transmission (b) spectra for a HLAMM with four curved channels connected at the center. The insets on the left indicate the normalized frequencies and pressure distributions for the artificial monopole and dipole Mie resonances. Omnidirectional band gaps are shaded in blue, while partial band gaps for the ΓX direction are depicted in green. The legend for the transmission spectrum indicates the number of unit cells used in the model.*

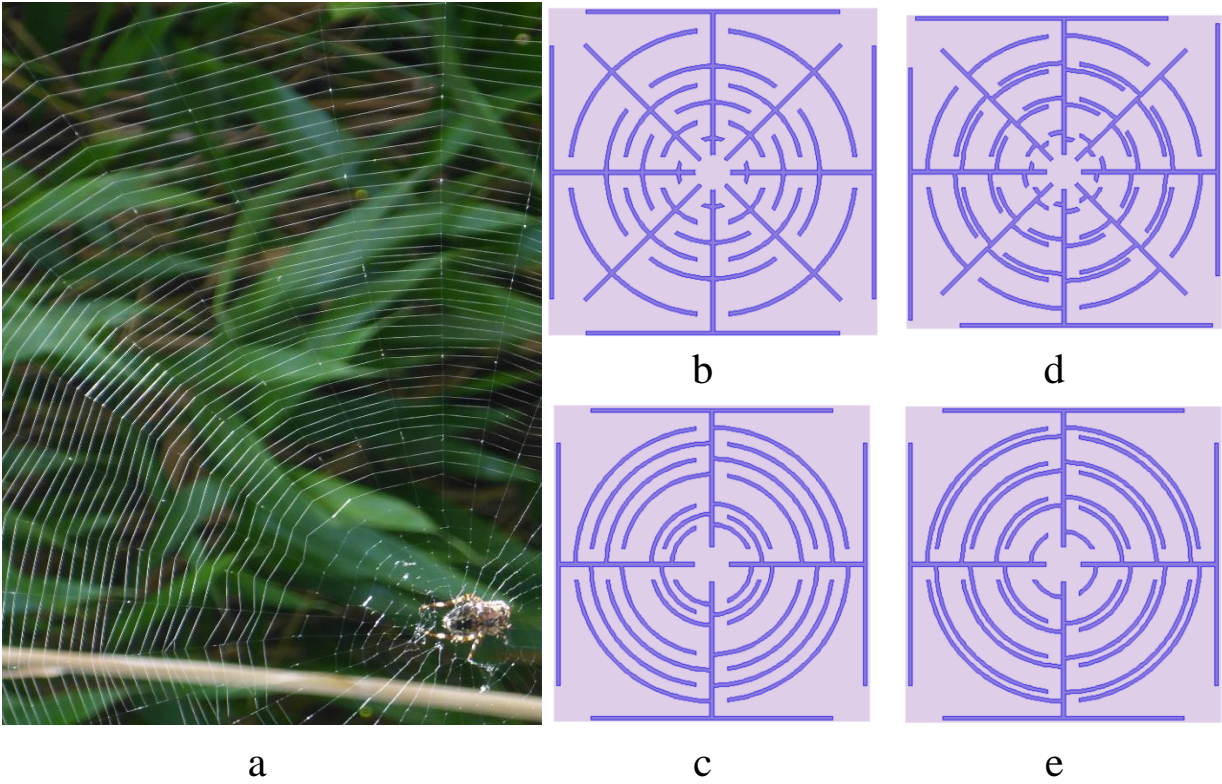

*Figure 5. (a) Spider web woven by a garden spider Araneus diadematus (photo by D. Krushinsky) (b) Unit cell of a labyrinthine metamaterial with a cavity between the curved channels after removing one of the solid walls from the initial geometry shown in Fid. 1a. (c) Unit cell with a cavity and two areas with curved channels of different width. This configuration resembles the architecture of the spider web in (a) and additionally allows to preserve approximately the same coiling coefficient as in Fig. 1b. (d-e) Unit cells with a periodically varying width of the curved channels.*

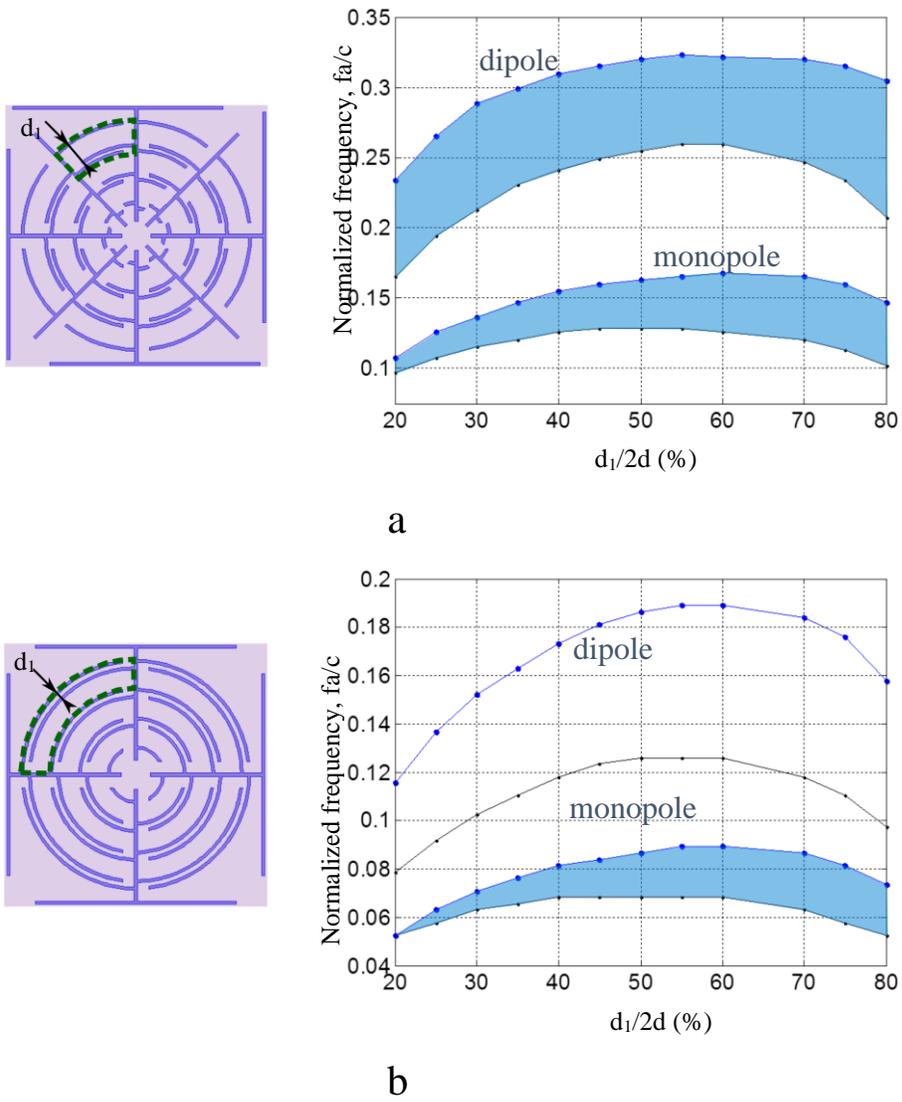

*Figure 6. Variation of the band gap frequencies (shaded in blue) and monopole and dipole Mie resonances for unit cells with a periodically varying width comprising (a) eight and (b) four curved channels.*

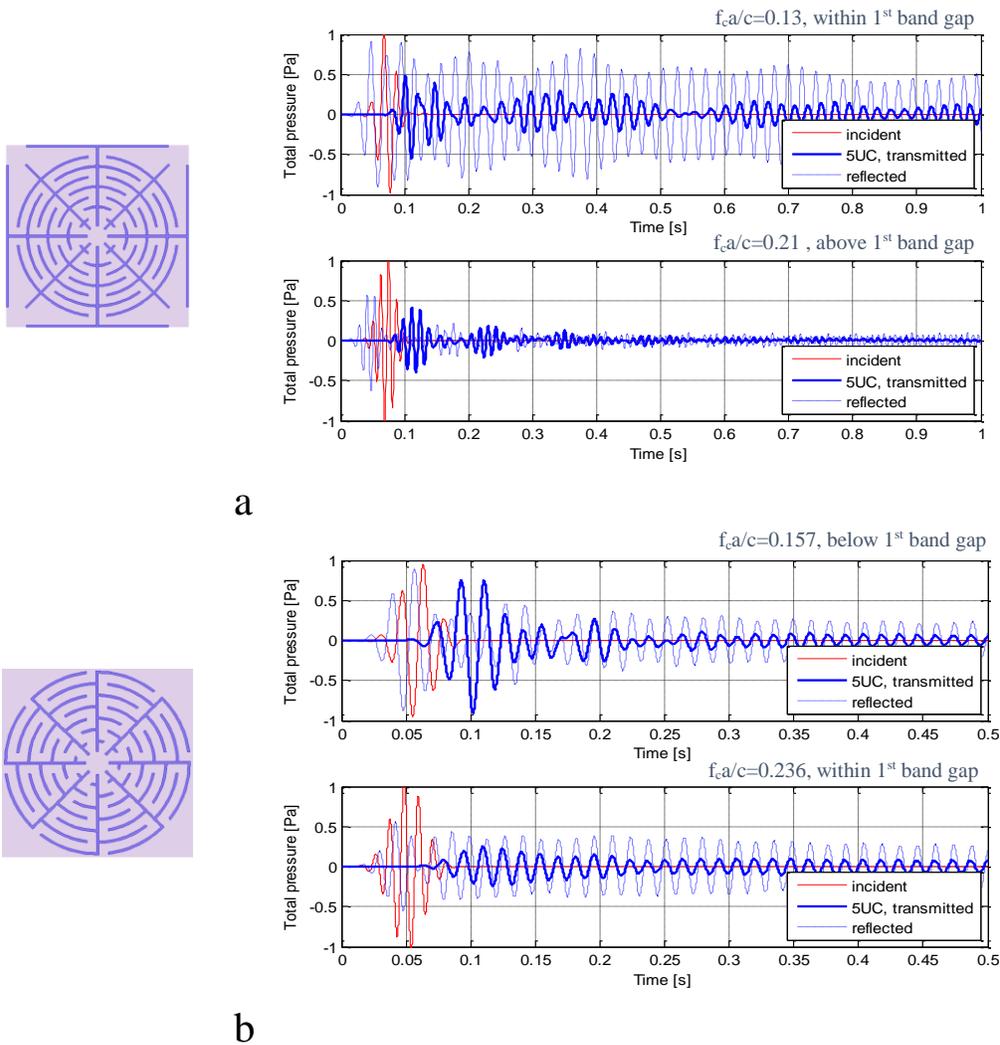

*Figure 7. Transient analysis of a Gaussian modulated sinusoidal pulse with the central frequency $f_c$ propagating though a set of 5 closely-located unit cells (shown in red). The presence of edge cavities (a) enables better wave reflection performance on band gap frequencies (governed by the monopole Mie resonance) compared to the metastructure without edge cavities (b).*

**Tables**

*Table 1. External radii $r_i$ $(i = 1, ..., 7)$ of solid walls for the spider web-inspired unit cells with eight curved channels as schematically shown in Figure 5c.*

|        | $r_1$ | $r_2$ | $r_3$ | $r_4$  | $r_5$ | $r_6$   | $r_7$ |
|--------|-------|-------|-------|--------|-------|---------|-------|
| Case 1 | 0.12a | 0.15a | 0.18a | 0.265a | 0.31a | 0.355a  | 0.4a  |
| Case 2 | 0.12a | 0.14a | 0.16a | 0.25a  | 0.3a  | 0.35a   | 0.4a  |
| Case 3 | 0.12a | 0.14a | 0.16a | 0.295a | 0.33a | 0.365a  | 0.4a  |